\documentclass[aps,a4paper]{revtex4}
\usepackage{epsfig}
\usepackage{graphicx}
\usepackage{amsmath,amssymb,color}
\usepackage[english]{babel}

\parskip=\medskipamount

\newcommand{\eq}[1]{(\ref{#1})}
\newcommand{\fig}[1]{Fig.~\ref{#1}}

\newcommand{\be}{\begin{equation}}
\newcommand{\ee}{\end{equation}}

\newcommand\disp{\displaystyle}

\newcommand{\la}{\left<}
\newcommand{\ra}{\right>}

\newcommand{\ve}{\mathbf}

\begin{document}

\title{Effective Hamiltonian of topologically stabilized polymer states}

\author{K. Polovnikov$^{1,2}$, S. Nechaev$^{3,4}$ and M.V. Tamm$^{2,5}$}

\affiliation{$^1$ Skolkovo Institute of Science and Technology, 143026 Skolkovo, Russia \\ $^2$ Faculty of Physics, Lomonosov Moscow State University, 119992 Moscow, Russia \\ $^3$Interdisciplinary Scientific Center Poncelet (ISCP), 119002, Moscow, Russia \\ $^4$Lebedev Physical Institute RAS, 119991, Moscow, Russia \\ $^5$ Department of Applied Mathematics, MIEM, National Research University Higher School of Economics, 101000, Moscow, Russia}

\date{\today}

\begin{abstract}
Topologically stabilized polymer conformations observed in melts of nonconcatenated polymer rings and crumpled globules, are considered to be a good candidate for the description of the spatial structure of mitotic chromosomes. Despite significant efforts, the microscopic Hamiltonian capable of describing such systems, remains yet inaccessible. In this paper we consider a Gaussian network -- a system with a simple Hamiltonian quadratic in all coordinates -- and show that by tuning interactions, one can obtain fractal equilibrium conformations with any fractal dimension between 2 (ideal polymer chain) and 3 (crumpled globule). Monomer-to-monomer distances in topologically stabilized states, according to our analysis of available numerical data, fit very well the Gaussian distribution, giving an additional argument in support of the quadratic Hamiltonian model. Mathematically, the resulting polymer conformations can be mapped onto the trajectories of a subdiffusive fractal Brownian particle. As a by-product of our study, two novel continual integral representations of the fractal Brownian motion are proposed.
\end{abstract}

\maketitle

\section{Introduction}

Classical statistical physics of polymers relies on the study of three archetypical polymer states: the ideal, swollen, and collapsed polymer chains \cite{deGennes_book,DoiEdwards, GrosbergKhokhlov, Rubinstein}. Equilibrium conformational statistics of linear polymers can be described by combinations of these models for any concentrations and chain interaction parameters.

In ideal macromolecules the elementary units do not interact with each other apart from being sequentially connected. Statistical description of ideal chains is based on the analogy between the equilibrium ensemble of ideal polymer chain conformations and trajectories of Brownian particles: similarly to the ensemble of random walks, ideal linear polymers in a free space have Gaussian statistics with the fractal dimension $d_f=2$. This analogy can be easily generalized to the case of ideal polymers in external potentials.

Swollen polymer state emerges due to the presence of so-called ``excluded volume interactions", i.e. repulsion between monomer units, which are distant along the chain but close in the space. The corresponding partition function can be interpreted as a self-avoiding random walk. The properties of swollen polymers are well understood due to the famous polymer-magnetic analogy discovered by de Gennes \cite{deGennes_pm} for solitary chains and extended by Des Cloizeaux \cite{desCloizeaux} to polymers in solutions. In particular, statistics of swollen chains in two- and three- dimensional spaces is known to be non-Gaussian, though self-similar, with corresponding fractal dimensions being equal to 4/3 in 2D and approximately $1.7$ in 3D.

Properties of collapsed polymer chains are governed by an interplay of attractive and repulsive interactions between monomer units. Implying the existence of attractive interactions only, one arrives at the unphysical conclusion that a polymer collapses into a point. Stabilization of a polymer chain in the collapsed regime is due to the equilibration between two-body attractive and three-body repulsive interactions. In the mean-field approximation the statistics of resulting states can be described in terms of an ideal chain in an external self-consistent field created by volume interactions among distant parts of the chain (or other chains in a multi-chain setting) \cite{Flory_book, LGK_review, deGennes_book}.

It has been becoming clear in recent years that these three classical archetypes do not exhaust the variety of macromolecular states existing in bio- and synthetic polymers. In particular, the statistics of ring polymers with fixed topology is definitely not covered by any of them. Contrary to linear polymers, rings preserve their topology: for example, initially nonconcatenated rings cannot get into a concatenated state without being ruptured. The resulting topological repulsion between nonconcatenated rings drastically changes statistical properties of chains in a melt \cite{nechaev,cates,sakaue,obukhov,grosberg14,ge16,everaers_grosberg}. It has been conjectured in \cite{nechaev} that conformations of long unknotted and non-concatenated ring polymers in melts are compact fractals with the fractal dimension $d_f=3$ starting from some minimal scale, called entanglement length $N_e$. This conjecture is now well-established both numerically (see, e.g. \cite{halverson_prl,grosberg_review}) and in several competing semi-analytical theories \cite{sakaue, obukhov, grosberg14,ge16}.

Contrary to ideal and swollen chains, the interactions in topologically stabilized globular polymers are substantially non-local. Moreover, in a dense system, such as a collapsed ring, the topology is not screened and an explicit microscopic Hamiltonian for non-phantom rings is unknown. Development of description of topologically interacting polymers from first principles remains an open fundamental problem. Interest to the topologically regulated polymer conformations is driven by experimental and numerical evidence that similar states may be observed as transient metastable conformations of linear polymers \cite{nechaev88,grosberg93} relevant for the understanding chromosome packing in living cells \cite{lieberman-eiden,mirny11}. This conjecture is based on the estimates that the lifetime of such transient states may exceed the biologically relevant timescales \cite{rosa08} (see also \cite{grosberg_review}). As an alternative to this view, there have been recently proposed several other possible models explaining chromosome packing in living cells. Some of them involve the concept of reversible bridging between parts of the chromosomes \cite{nicodemi} and non-equilibrium loop extrusion processes \cite{loop_extrusion, loop_extrusion2}. All these models have a common feature: in a wide range of length scales, the resulting equilibrium chromatin packing is fractal with the fractal dimension $d_f$ lying in the interval $2\leq d_f \leq 3$. However, the microscopic Hamiltonian of these self-similar conformations is unknown, which sufficiently hardens the analytical tractability of corresponding theories.

In this paper we show that it is possible to design a Hamiltonian of volume interactions for a polymer chains in such a way that the resulting polymer conformations in thermal equilibrium are fractal with prescribed fractal dimension $2\leq d_f\leq 3$. The statistics of resulting chain conformations is identical to the statistics of trajectories of a fractal Brownian motion (fBm) \cite{mandelbrot}. In this sense, our result is a generalization of the classical analogy between Brownian motion and ideal polymer chain.

The paper is organized as follows. In Section II we recall a mapping of polymer conformations onto particle trajectories. In Section III we provide the microscopic Hamiltonian generating Gaussian polymer conformations and prove that such a description is identical to the theory of the fractal Brownian motion. In Section IV we generalize the memory-dependent action derived in \cite{oshanin} and establish its connection with the action of a fBm particle. In Section V we show that the simulation data from earlier works \cite{imakaev14,tamm15}, where topologically stabilized polymer states were simulated, are consistent with the Gaussian monomer-to-monomer distribution typical of the quadratic Hamiltonian introduced in this paper, which makes us believe that our proposed Hamiltonian is a good candidate for phenomenological description of these states.

\section{Fractal Brownian motion as a conformation of a polymer chain}

Statistical properties of long ($N \gg 1$) polymer chains are insensitive to specific microscopic details of chain flexibility, which leaves us a freedom to choose a particular microscopic model of a chain. Here we use a beads-on-string model of a polymer chain with pairwise interactions between the beads. The chain conformation is characterized by coordinates of all $N+1$ units, $\ve X = \{\ve x_0, \ve x_1, ..., \ve x_N\}$. The typical bead-to-bead distance is a fluctuating variable with the mean square $a^2$, so the total length of the chain is $L = Na$. The potential energy $U(\ve X)$ of volume interactions between the beads is assumed to be a sum of pairwise interactions $V(\ve x_n, \ve x_{n'})$. The partition function $P(\ve x_k, \ve x_m)$ of the chain with $k$-th and $m$-th beads fixed at $\ve x_k$ and $\ve x_m$, respectively, can be expressed in terms of the Euclidean Feynman path integral (the Wiener measure in the probabilistic language) \cite{DoiEdwards} with the action:
\be
P(\ve x_k, \ve x_m) =\int {\cal D}\{\mathbf{X}\}\,e^{-S\{\ve X \}}; \qquad S = \frac{3}{2 a^2} \sum_{n=0}^{N-1}\left(\ve x_{n+1} - \ve x_n\right)^2 + \sum_{n=0}^{N}\sum_{n'=n+2}^{N} V\left(\ve x_n, \ve x_{n'}\right),
\label{path2}
\ee
where the integration is taken over all possible conformations, ${\cal D}\{\mathbf{X}\} = \prod_{n \ne k, m} d \ve x_n$. Here and below we measure all energetic terms in the dimensionless units or equivalently $k_B T = 1$.

In the absence of volume interactions the partition function \eq{path2} obeys the diffusion equation with $\ve x=\ve x_k - \ve x_m$ and $s = |k-m|$ playing roles of coordinate and time, respectively. Therefore, equilibrium distribution of the monomer-to-monomer distance is the same as for the standard Brownian motion:
\be
P({\ve x}_k, {\ve x}_m, s) = \left(\frac{3}{2\pi s a^2} \right)^{3/2} \exp\left(-\frac{3 ({\ve x}_k - {\ve x}_m)^2}{2 s a^2}\right).
\label{gaussian}
\ee
The distribution \eq{gaussian} means that conformations of ideal polymer chains are fractals with $d_f = 2$ similarly to Brownian trajectories. Here we generalize this analogy to the case of arbitrary fractal dimension $d_f$. Namely, we ask whether it is possible to choose pairwise interactions $V\big(\ve x_n, \ve x_n'\big)$ in \eq{path2} in such a way that the resulting equilibrium monomer-to-monomer distances would have a Gaussian distribution with some prescribed fractal dimension $d_f$:
\be
P({\ve x}_k, \ve x_m, s) = \left(\frac{3}{2\pi a^2 s^{2/d_f}}\right)^{3/2} \exp\left(-\frac{3 ({\ve x}_k - \ve x_m)^2}{2 a^2 s^{2/d_f}}\right)
\label{fGaussian}
\ee

The behavior dictated by \eq{fGaussian} is typical for the fBm, $B_H$, with $H = 1/d_f$, a process whose increments are the integrals over increments of ordinary Brownian motion weighted with a non-local algebraic memory kernel \cite{mandelbrot}. This process is strongly non-Markovian in a sense that correlations of fBm increments (positive for $H>1/2$ and negative for $H<1/2$) decay as a power-law. However, fBm is a linear function of Brownian motion, and is Gaussian in sense of \eq{fGaussian}. It is, therefore, an example of a Gaussian process with a scale-free memory. Importantly, fBm has stationary and self-similar increments. This makes it a plausible candidate for the description of crumpled polymer conformations.

There are several ways of constructing a Langevin formalism, which generate a process with fBm statistics. However, if one adds a requirement that the resulting process should also respect the fluctuation-dissipation theorem, there is a preferred form, known as fractional Langevin equation (fLe) in the overdamped limit \cite{kubo,hanggi,deng}:
\be
\xi_H \int_{0}^{t}d\tau K(t-\tau)\frac{d\mathbf{r}(\tau)}{d\tau} = \mathbf{F}_H(t); \qquad \langle \mathbf{F}_H(t_1) \mathbf{F}_H(t_2) \rangle = \xi_H K(t_1 - t_2); \qquad K(t_1 - t_2) = \frac{2(1-H)(1-2H)}{|t_1-t_2|^{2H}}.
\label{fle}
\ee

In this paper, we show that for long polymer chains ($N \gg 1$) a pairwise potential:
\be
V(\ve x_k, \ve x_{m}) = a_{km} \left(\mathbf{x}_{k}-\mathbf{x}_{m}\right)^2
\label{quadr_potential}
\ee
can be used to construct polymer chains with fBm-like equilibrium distribution of the monomer-to-monomer distance \eq{fGaussian} with any $2 < d_f < 3$ provided that coefficients $A_{km}$ depend only on chemical distance between monomers $|k-m|=s$ and decay asymptotically at $s\gg 1$ as
\be
a_{s} \sim \disp c\, s^{-\gamma}; \quad \gamma\in (2,3)
\label{decay}
\ee
with $c>0$. The resulting large-scale fractal dimension of conformational statistics is related to the decay exponent $\gamma$ by
\be
d_f = \frac{2}{\gamma-2}
\label{df}
\ee
If the coefficients in \eq{quadr_potential} decay faster than $s^{-3}$, the statistics of the corresponding polymer chain remains ideal at large scales and the monomer-to-monomer distance is given by \eq{gaussian}. The value of $\gamma = 3$ is critical, giving rise to logarithmic corrections in \eq{gaussian}.

\begin{figure}[ht]
\centerline{\includegraphics[width=8cm]{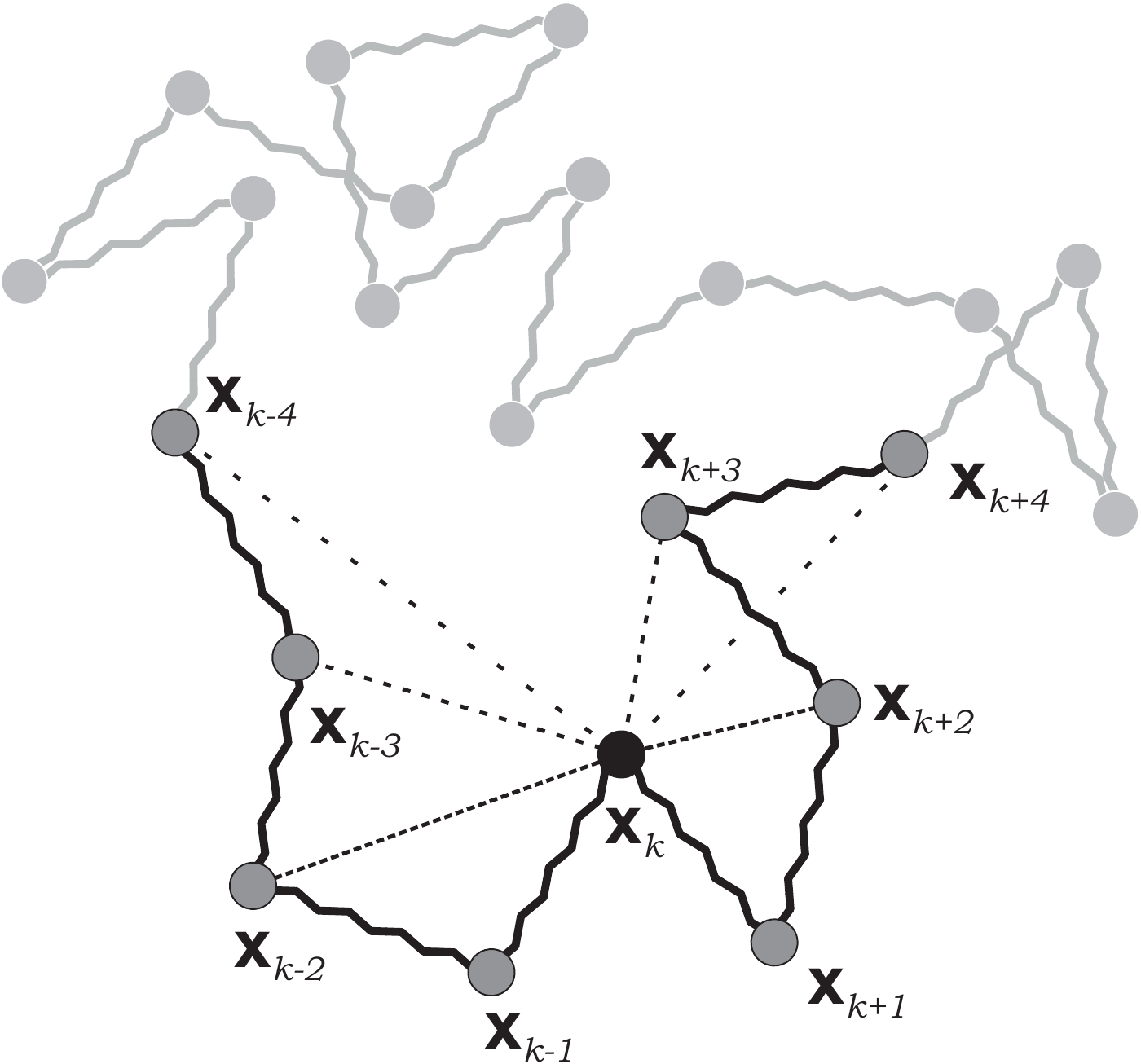}}
\caption{Schematic image of the pairwise interactions \eq{quadr_potential} $V_{\ve x_k, \ve x_{m}}$ of the $k$-th monomer ($\mathbf{x}_k$) with adjacent monomers of the chain with coordinates $\mathbf{x}_{k\pm 1},\mathbf{x}_{k\pm 2},\mathbf{x}_{k\pm 3},...$. Elastic constants $a_{km}$ decay algebraically which is depicted by dashed lines with increasing spacing.}
\label{fig:01}
\end{figure}

Quadratic interactions in \eq{quadr_potential} can be interpreted as a set of strings of varying rigidity connecting each pair of monomers, as shown in \fig{fig:01}. It makes sense, therefore, to incorporate the nearest-neighboring harmonic interactions (i.e., the first term in the action \eq{path2}) into the definition of $V$, such that $a_{k,k+1}=3/(2a^2)$. The potential of the form \eq{quadr_potential} has been studied previously in various contexts. In particular, the resulting Gaussian networks \cite{bahar97,haliloglu97,min} with $(m,k)$-depending rigidities are often used for the description of 3D structures of proteins. In \cite{dolgushev1,dolgushev2} static and dynamic properties of marginally compact trees with various fractal architectures were considered. A related hierarchical variational approach for an account of volume interactions of swollen polymer chains had been proposed in Ref. \cite{burlatsky}. In a dynamic context, the "beta-model" \cite{holcman}, which is a Rouse-like model of a polymer chain with a time relaxation spectrum of a certain specific form. In Ref. \cite{polovnikov18} a similar model was used for studying dynamic properties of a crumpled globule in a viscoelastic environment.

On the base of a quadratic potential \eq{quadr_potential}, we propose an alternative form of the action generating long fractal Brownian conformations, $N \to \infty$, which consists in the modification of the "kinetic" term in \eq{path2}:
\be
\tilde{P}(\mathbf{x},\mathbf{y}) = \int {\cal D}\{\mathbf{X}\}\,e^{\tilde{S}}; \qquad \tilde{S} = \int_0^\infty d\xi \int_0^\infty d\xi' \; \frac{\partial \mathbf{X}(\xi)}{\partial \xi} \frac{\partial \mathbf{X}(\xi')}{\partial \xi'}\varphi(|\xi-\xi'|)
\label{path}
\ee
where the function $\varphi(|\xi-\xi'|)$ is a power-law decaying memory kernel. Clearly, $\varphi(\xi)=\delta_{\xi}$ corresponds to a simple Brownian motion with $H=1/2$. Action of the form \eq{path} has been previously appeared in \cite{oshanin}, where it was shown that \eq{path} with $\varphi(\xi) =\xi^{-1/2}$ corresponds to the statistics of trajectories of the Rouse particles, which is known to be fBm with $H=1/4$. Here we generalize this result and show that for any $0<H<1/2$ the corresponding ensemble of fractal Brownian trajectories can be obtained from the action \eq{path} with $\varphi(s) \sim s^{-2H}$.

\section{Gaussian chain with long-range quadratic interactions}

Here we prove the results outlined above, which connect the modes described by \eq{quadr_potential}, and \eq{decay} with fractal Brownian motion behavior \eq{fGaussian}. To simplify the consideration, consider a ring chain of $N \gg 1$ monomers, $\ve x_N \equiv \ve x_0$. We consider phantom chains here, so for $|k-m| \ll N$ the distribution of $\ve x_k - \ve x_m$ does not depend on boundary conditions and this assumption does not lead to any loss of generality. Also, assume for definiteness that $N$ is odd, $N=2n+1$. The potential \eq{quadr_potential} in this case takes the following form
\be
V (\{\ve X\}) = \psi (\ve x_0, \ve x_1, ..., \ve x_{N-1}) = \sum_{m<k} a_{s}(\mathbf{x}_k-\mathbf{x}_m)^2
\label{quadr_potential_2}
\ee
where $s = s(k,m)$ is the shortest contour distance between monomers $k$ and $m$:
\be
s(k,m) = \min \left( |k-m|, N-|k-m| \right)
\ee
This distance $s(k,m)$ is a symmetric and circularly periodic function
\be
s(k,m) = s(m,k); \; s\left(k+i \mod N,\,\,m+i \mod N\right) = s(k,m)
\ee
It means that there are $n$ independent different values of $a(s), \; s=1..n$. Introducing $a(0) = 2\sum_{s>0} a(s)$ one can rewrite potential \eq{quadr_potential_2}
\be
V (\{\ve X\}) = \psi (\ve x_0, \ve x_1, ..., \ve x_{N-1}) = a(0) \sum_{m=0}^{N-1} x_m^2 - 2 \sum_{m<k} a(s(k,m)) x_m x_k = \langle \ve X| \mathbb A | \ve X \rangle
\label{quadr_a}
\ee
where the matrix $\mathbb A$ is Laplacian and circulant. Its eigenvectors $\ve A_p$, ${\mathbb A}|\ve A_p \rangle= \omega_p |\ve A_p \rangle$, have coordinates
\be
\ve A_p^{(k)} = \frac{1}{\sqrt{N}}\exp \left( \frac{2\pi i p k}{N}\right); \quad k = 0, 1 ... N-1
\label{eigc}
\ee
and the eigenvalues $\omega_p$ are
\be
\omega_p = a(0) - \sum_{s=1}^{n} a(s) \left(\exp \left( \frac{2\pi i p s}{N}\right) + \exp \left( \frac{2\pi i p (N-s)}{N}\right)\right) = 2 \sum_{s=1}^{n} a(s) \left(1- \cos \left( \frac{2\pi p s}{N}\right) \right)
% = \hat{\mathbb C}_p\{a_s^{-1}\}
\label{eigv}
\ee
Importantly, $\omega_0 = 0$ has degeneracy 2 (as it should be for a Laplacian matrix and other eigenvalues): $\omega_p = \omega_{N-p}$. Moreover, the spring constants $a(s)$ should decay faster than $1/s$ in order for expressions in \eq{eigv} to converge. Physically, it means that strongly attractive elastic networks with slower decay of $a_s$ get collapsed into a single point in the limit $N \to \infty$. In the Appendix A we consider a particular case of $a(s)$ decaying as a general power law $a(s) = c s^{-\gamma}$ and show (see \eq{a6}) that in this case the eigenvalues with $p \ll N$ behave as
\be
\omega_p \sim \left\{ \begin{array}{cl} \disp & \Gamma (1-\gamma) \left(\frac{ p}{N}\right)^{\gamma-1} \; \; \mbox{for $2 < \gamma < 3$} \medskip \\
\disp &\frac{1}{\gamma-3}\left(\frac{p}{N}\right)^{2} \;\; \mbox{for $\gamma > 3$}
\end{array} \right. ,
\label{omega_spectrum}
\ee
where we keep only coefficients divergent at $\gamma \to 3$.

The equilibrium properties of an elastic network are easier to analyze in terms of normal relaxation modes, $\ve u_p = \left\langle\ve X | \ve A_p\right\rangle$, $p=0..N-1$
\be
\begin{array}{rll}
\ve u_p & = & \dfrac{1}{\sqrt{N}}\displaystyle \sum_{k=0}^{N-1} \ve x_n \exp \left(\dfrac{2\pi ipk}{N}\right), \medskip \\
\quad \ve x_k &= &\dfrac{1}{\sqrt{N}}\displaystyle \sum_{p=0}^{N-1} \ve u_p \exp \left(- \dfrac{2\pi ipk}{N}\right)
\end{array}
\label{fourier}
\ee
In the new coordinates the potential \eq{quadr_a} can be diagonalized, providing the following form
\be
V (\{\ve X\}) = \sum_{m=0}^{N-1} \langle \ve A_m |\ve u_m^{*} \sum_{p=0}^{N-1} \omega_p \ve u_p |\ve A_p \rangle = \sum_{m=0}^{N-1} \sum_{p=0}^{N-1} \ve u_m^{*} \omega_p \ve u_p \langle \ve A_m |\ve A_p \rangle = \sum_{p=0}^{N-1} \omega_p \left|\ve u_p\right|^2
\label{diag_potential}
\ee
where in the last equation we used that $\langle \ve A_m| \ve A_p \rangle =\delta_{m,p}$. In equilibrium, the distribution of energy between the addenda of \eq{diag_potential} obeys the equipartition theorem, and therefore
\be
\overline{ \ve u_p^* \ve u_{p'}} = \frac{3 \delta_{p p'}}{\omega_p}
\label{equipartition}
\ee
where the bar denotes the equilibrium ensemble averaging.

Now, to prove that in the equilibrium the monomer-to-monomer distance $\ve x_k - \ve x_m$ for $1\ll s(k,m) \ll N$ is given by the fBm distribution \eq{fGaussian} we need to prove two statements: (i) that the equilibrium distribution is Gaussian, and (ii) that its variance grows as a power of $s$.

The statement (i) follows straightforwardly from the fact that the statistical weight of the full conformation $\ve X = \{ \ve x_0, .., \ve x_{N-1}\}$ is a Gaussian function:
\be
P_N(\ve X) = \frac{1}{Z_N}\exp\left\{-\langle \ve X| \mathbb A | \ve X \rangle \right\}.
\label{path3}
\ee
where $Z_N$ is the partition function, and the Hamiltonian is given by \eq{quadr_a}. Since the Hamiltonian is translationally invariant, we get:
\be
P(\mathbf{x}_{k}, \mathbf{x}_{m}, s) = \frac{1}{Z_N} \int \exp\left\{- \langle \ve X| \mathbb A | \ve X \rangle \right\} \prod_{i \neq k,m} d\ve x_i = \left(\frac{3}{2\pi \sigma_{km}}\right)^{3/2} \exp\left(-\frac{(\mathbf{x}_k-\mathbf{x}_m)^2}{2\sigma_{km}^2}\right)
\label{two-point-distr}
\ee
where the variance $\sigma_{km}^2 = \overline{\left(\ve x_{k} - \ve x_{m}\right)^2} \equiv \sigma^2(s)$, which is some function of the contour length $s(k,m)$. Rewriting this variance in terms of the normal modes \eq{fourier} one gets:
\be
\sigma^2 (s) = \frac{1}{N} \overline{\left|\sum_{p=0}^{N-1} \ve u_p \left(e^{\frac{-2\pi ipk}{N}}-e^{\frac{-2\pi ipm}{N}}\right)\right|^2} = \frac{12 k_B T}{N} \sum_{p=1}^{n} \omega_p^{-1} \left(1 - \cos\left(\frac{2\pi p s(k,m)}{N}\right)\right),
\label{sigma}
\ee
In \eq{sigma} we used the degeneracy of the spectrum and the equipartition theorem \eq{equipartition}.

\begin{figure}[ht]
\centerline{\includegraphics[width=16cm]{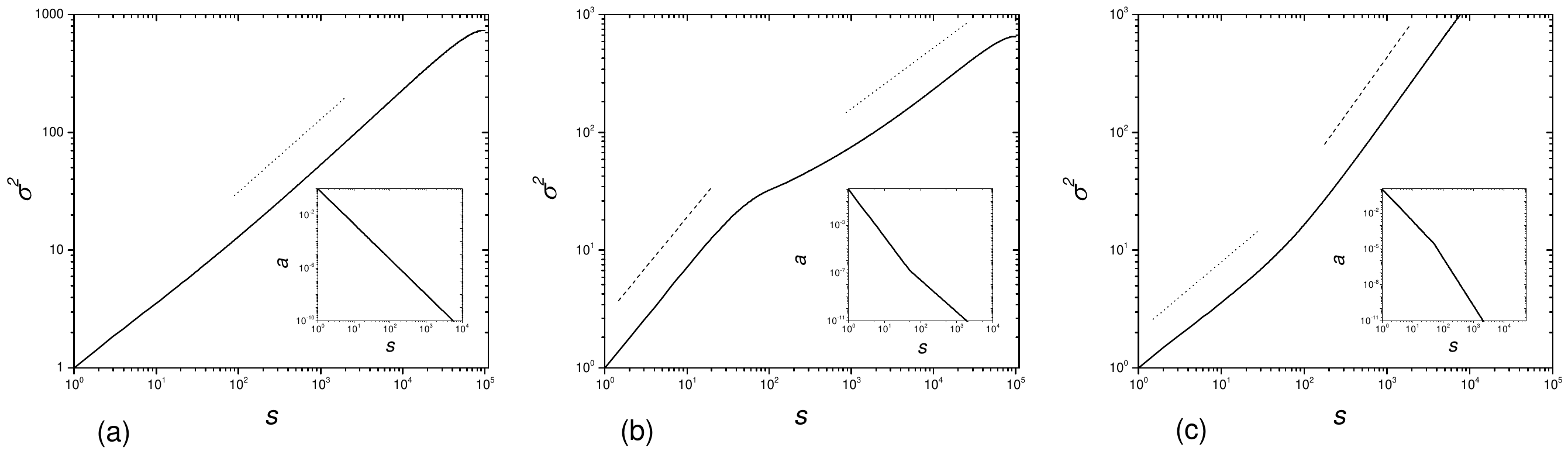}}
\caption{Behavior of the dispersion $\sigma^2(s), s = 1,...,n$ defined in \eq{sigma} for a ring chain with $n = 10^5$ for three cases: (a) single power-low decay $a(s) = s^{-8/3}$; (b), (c) combination of two power laws \eq{combined} (b): $\gamma_1 = 4$, $\gamma_2 = 8/3$, $s^* = 100$, (c): $\gamma_1 = 8/3$, $\gamma_2 = 4$, $s^* = 100$. Dash lines correspond to $\sigma^2(s) \sim s^{2/3}$ (short dashes) and $\sigma^2(s) \sim s$ (long dashes). The plots are rescaled so that $\sigma^2(s)=1$ for $s = 1$.}
\label{fig:02}
\end{figure}

To prove the statement (ii), note that the asymptotic behavior of \eq{sigma} for $s \gg 1$ is controlled by the behavior of $\omega_p$ for $p \ll N$ and the typical relevant $p$ is of order $N/s$. Therefore, to have algebraically decaying coefficients $a(s)$, one can use the expression \eq{omega_spectrum}, which gives
\be
\sigma^2 (s) \sim \frac{1}{N} \sum_{p=1}^{n} \left(\frac{ p}{N}\right)^{1- \bar{\gamma}} \left(1 - \cos\left(\frac{2\pi p s(k,m)}{N}\right)\right) \sim \int_0^{\pi} x^{1- \bar{\gamma}} (1-\cos xs) dx = s^{\bar{\gamma}-2}\int_0^{\pi s} y^{1- \bar{\gamma}} (1-\cos y) dy,
\label{sigma2}
\ee
where we used the notation
\be
\bar{\gamma} = \left\{ \begin{array}{cl} \disp & \gamma \; \; \mbox{for $2 < \gamma < 3$} \medskip \\ \disp &3 \;\; \mbox{for $\gamma > 3$}
\end{array} \right.
\label{r}
\ee
The integral in the right hand side converges for all relevant $\bar{\gamma}$ and for $s\gg 1$ only weakly depends on its upper limit, which allows us to extract the leading asymptotic
\be
\sigma_s^2 \sim \left\{ \begin{array}{cl} \disp &s^{\gamma-2} \; \; \mbox{for $2 < \gamma < 3$} \medskip \\
\disp &s \;\; \mbox{for $\gamma > 3$}
\end{array} \right.
\label{sfinal}
\ee
Thus, if $a(s)$ decays slower than $s^{-3}$, the equilibrium conformations have fractal dimension $d_f = 2/(\gamma-2)$, while for faster decays of $a(s)$ the chain adopts ideal conformation akin to the standard Brownian trajectory, and the presence of additional terms in the potential (additional harmonic springs between beads) just renormalizes the chain stiffness. Equilibrium conformation of a chain is, therefore, an fBm with the Hurst exponent
\be
H = \frac{\bar{\gamma}}{2}-1 = \left\{ \begin{array}{cl} \disp &\gamma/2 - 1 \; \; \mbox{for $2 < \gamma < 3$} \medskip \\ \disp &1/2 \;\; \mbox{for $\gamma > 3$}
\end{array} \right.
\label{Hgamma}
\ee
This result is, so far, obtained just for the case when $a(s)$ decay strictly as a power law. In order to address a general situation, we evaluated \eq{eigv} and \eq{sigma} numerically for several specific choices of $a(s)$ in particular, of the form
\be
a(s) = \left\{ \begin{array}{cl} \disp & c_1 s^{-\gamma_1}\; \; \mbox{for $s<s^*$} \medskip \\ \disp &c_2 s^{-\gamma_2}, \quad \mbox{for $s>s^*$}
\end{array} \right. ,
\label{combined}
\ee
The corresponding behavior is shown in \fig{fig:02}. We see that in this case the chain as a whole is not a fractal anymore. Separation of scales is clearly seen: for $s \ll s^*$ the behavior of $\sigma^2$ is controlled by the exponent $\gamma_1$ while for $s\gg s^*$ it is controlled by $\gamma_2$. This means that not only the large-scale behavior of $\sigma^2$ depends only on large-scale behavior of $a(s)$ in agreement with \eq{sfinal}, but also that one can use the Hamiltonian \eq{quadr_a} to construct polymer conformations with different fractal dimensions on different length scales and/or for different parts of the chain. This might be useful, e.g., for the description of heterochromatin consisting of active and inactive domains (see. e.g., \cite{jost14,nazarov15,ulianov16}).

Interestingly, the interpretation of a fBm trajectory as a specific type of polymer conformation suggests a natural way to determine the power spectrum $f(p)$ of the fBm. From point of view of the polymer analogy $f(p)$ is related to the energy stored in the $p$-th normal mode, so equipartition theorem connects it with the eigenvalues $\omega_p$ of the interaction matrix $\mathbb A$:
\be
f(p) = \overline { \ve u_p^* \ve u_p} \sim \omega_p^{-1}.
\label{eqt}
\ee
Taking into account \eq{omega_spectrum} and \eq{Hgamma} one gets
\be
f(p) \sim \left(\frac{N}{p}\right)^{2H+1}.
\label{fgamma}
\ee
which is a known result for the fBm \cite{reed95}. Interestingly, within the Rouse approach to polymer dynamics, which corresponds to postulating
\be
\frac{\partial \ve x_i}{\partial t} = \frac{\partial V (\ve X)}{\partial \ve x_i} + \delta\text{-correlated Gaussian noise}
\ee
as equations of motion for individual monomers, $f(p)$ is also proportional to the relaxation time $\tau_p$ of the $p$-th mode \cite{tamm15,polovnikov18}.

\section{Action with algebraically decaying memory kernel}

In this section we discuss how to reinterpret the quadratic Gaussian interactions with algebraically decaying coefficients as an action with a modified kinetic term as suggested by \eq{path}. The partition function of a polymer chain with quadratic interactions \eq{quadr_a} reads
\be
Z_N = \int {\cal D}\{\mathbf{X}\}\,e^{-S};
\label{path3a}
\ee
where $S$, in the sense of a moving particle, is the Euclidean action $S$ which coincides with Hamiltonian of the polymer chain
\be
S = \la \ve X| \mathbb A | \ve X \ra,
\label{path3}
\ee
and integration is taken over ${\cal D} \{\mathbf{X}\} = \prod_{k=0}^N d\ve x_k$.

Discretizing the memory-dependent action in \eq{path}, one gets:
\be
\begin{array}{rll}
\tilde{S} &=& \displaystyle \int_0^N d\xi \int_0^N d\xi' \dfrac{\partial \mathbf{X}(\xi)}{\partial \xi} \dfrac{\partial \mathbf{X}(\xi')}{\partial \xi'} \varphi(|\xi-\xi'|) \medskip \\ &\approx & \displaystyle\sum_{k,m}^N (\mathbf{x}_k - \mathbf{x}_{k-1})(\mathbf{x}_m - \mathbf{x}_{m-1}) \varphi_{k,m} \medskip \\ & = & \displaystyle \sum_{k,m}^N \left(\varphi_{k,m} - \varphi_{k,m+1} - \varphi_{k+1,m} + \varphi_{k+1,m+1} \right) \mathbf{x}_k \mathbf{x}_m
\end{array}
\label{sbeta1}
\ee
We see that indeed the two expressions \eq{path3} and \eq{sbeta1} are equal provided that
\be
a_{km} = - \left(\varphi_{k,m} - \varphi_{k,m+1} - \varphi_{k+1,m} + \varphi_{k+1,m+1} \right)
\ee
for all $k, m$. For $1\ll |k-m| <n$ this reduces to
\be
a(s=|k-m|) = \left(\varphi(s-1) + \varphi(s+1) - 2\varphi(s)\right) \simeq \frac{\partial^2 \varphi(s)}{\partial s^2}
\label{discrete}
\ee
As we have shown in the previous section, the large-scale statistics of the chain depends only on the asymptotic behavior of $a(s)$. Thus, it is insensitive to particular details of the behavior of $a(s)$ or $\varphi (s)$ at small $s$. Assuming that for $s \to \infty$
\be
a(s) \sim s^{-\gamma},\quad \gamma \in (2,3),
\label{asymp_a}
\ee
one can approximate the difference in \eq{discrete} by the continuous derivative. Thus, we arrive at the conclusion that \eq{asymp_a} is equivalent to
\be
\varphi(s) \simeq \dfrac{s^{2-\gamma}}{(\gamma-2)(\gamma-1)}, \quad \gamma \in (2,3)
\label{asymp_phi}
\ee
Combining \eq{asymp_phi} with \eq{Hgamma}, we see that any action of the form \eq{sbeta1} with $\varphi(s)$ decaying as $s^{-2H}$, where $H\in (0, 1/2)$ for large $s$, generates an equilibrium ensemble of trajectories which are asymptotically equivalent to the fractal Brownian motion with the Hurst exponent $H$. In particular, for $H = 1/4$ we recover the action generating Rouse trajectories \cite{oshanin}, while the case $H = 1/3$ corresponds to the Hurst exponent of the crumpled globule.

Interestingly, it is possible to link the discussed representation of the fBm action with the fractional Langevin equation (fLe) \eq{fle} via a fluctuation-dissipation argument. The left-handed side of \eq{fle} corresponds to a dissipative friction force $\mathbf{F}$ acting on a fLe particle. At equilibrium, the average energy of the particle is conserved and the work performed by this force should be equal to the integral of the action $\tilde{S}$ along the trajectory of the particle. For a particle moving from $\mathbf{x}_1$ to $\mathbf{x}_2$ during the time $t$, the equations \eq{sbeta1} -- \eq{asymp_phi} adopt the form:
\be
\tilde{S} = 2\int_{0}^t dt' \int_{0}^{t'} dt'' \frac{\partial \mathbf{x}(t')}{\partial t'} \frac{\partial \mathbf{x}(t'')}{\partial t''} \varphi(t' - t'') = -\int_{\mathbf{x}_1}^{\mathbf{x}_2} \mathbf{F} \; d\mathbf{x}
\label{fsh}
\ee
Differentiating \eq{fsh}, one gets the expression for the force $\mathbf{F}$ the following expression:
\be
\mathbf{F}(t) = -\int_0^t dt' K_\alpha (t-t') \frac{\partial \mathbf{x}(t')}{\partial t'}; \quad K_\alpha(t-t') \sim \frac{1}{|t-t'|^{2H}},
\label{fh}
\ee
which, up to the choice of numerical coefficients, is identical to the one in the right hand side of \eq{fle}.

Thus, the analogy between conformation of a polymer chain and trajectory of a subdiffusive fBm particle allows describing the latter in terms of an action that implies velocity-velocity correlations with algebraically decaying memory kernel. This action is can be used to calculate the work performed by the friction force along the trajectory, and the resulting friction coincides with that prescribed by fractional Langevin equation thus shedding some light on the physical basics behind this equation. Note that in equilibrium the energy loss due to friction is compensated on average by the action of a fractional noise in the thermostat, which in the formalism presented here emerges from the summation over ``ghost'' interactions between the particle velocity at a given point and velocities in all its future positions.

\section{Discussion}

In this paper we have shown that a polymer chain described by the Hamiltonian of the form \eq{quadr_a} with coefficients decaying algebraically at large separation distances $a_{km}\sim |k-m|^{-\gamma} \text{ for } |k-m|\gg 1$ adopts a fractal Gaussian conformation with monomer-to-monomer distances growing as $|k-m|^{1/2}$ for $\gamma>3$ and as $|k-m|^{(\gamma-2)/2}$ for $\gamma \in (2,3)$. Putting it in other words, this means that adjusting parameters in \eq{quadr_a} one can construct fractal \emph{Gaussian} polymer conformations with any fractal dimension $d_f \geq 2$.

How physically relevant is this result? Can one, for example, use this Hamiltonian as a proxy way to describe topologically stabilized polymer states? The answer depends, to a large extent, on whether these polymer states, like nonconcatenated rings in a melt and mitotic chromosomes, are Gaussian or not. If they are, the potential \eq{quadr_a} seems to be a good phenomenological Hamiltonian for such systems in the absence of an exact microscopic one, while if they are not, it can only be used to reproduce those properties of real chains which depend on fractal dimension only.

\begin{figure}[ht]
\centerline{\includegraphics[width=16cm]{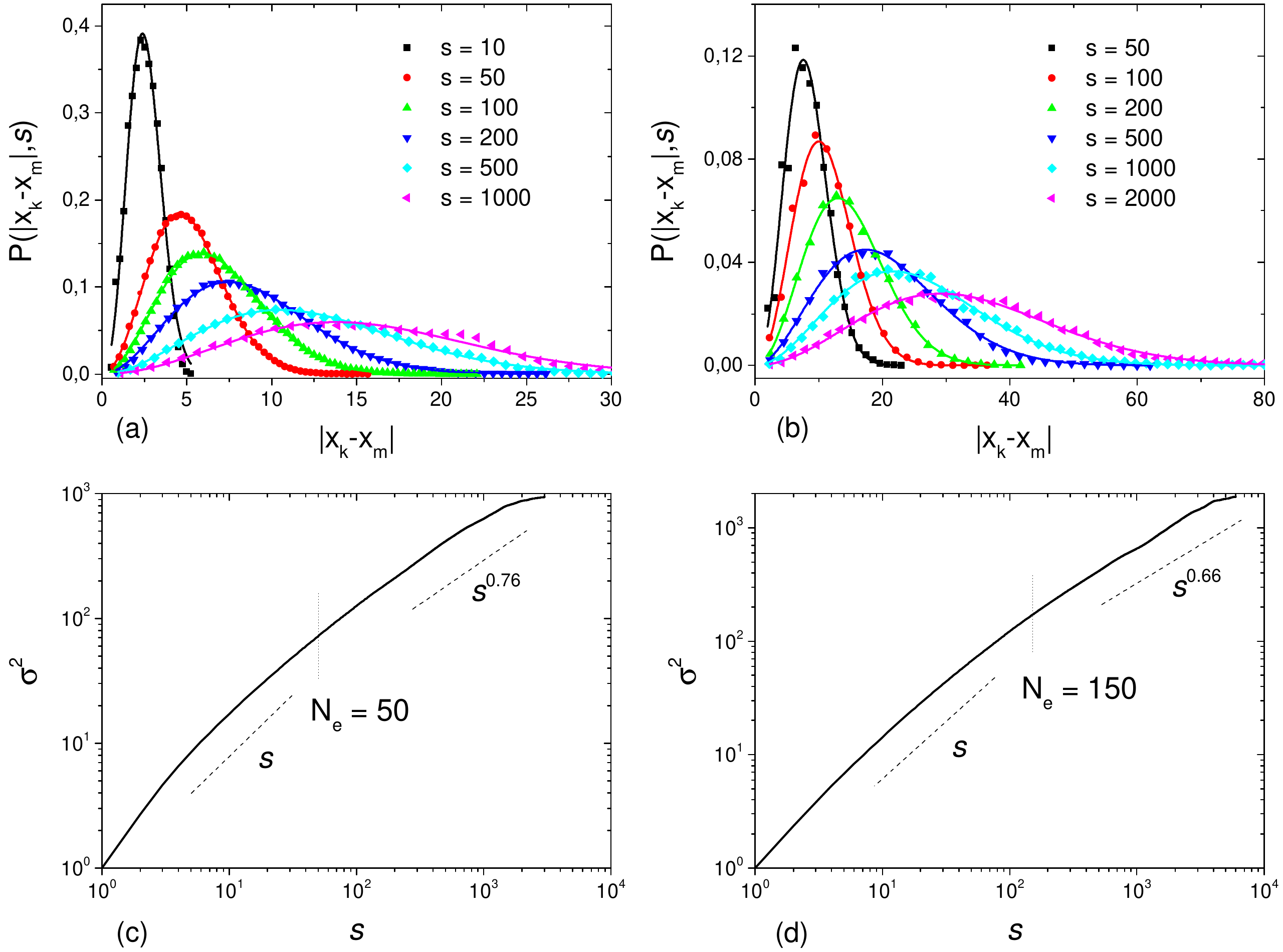}}
\caption{(a-b) Distribution of the monomer-to-monomer distance $P(|\mathbf{x}_k - \mathbf{x}_m|, s)$ for different $s$ for (a) partially equilibrated unknoted linear chain simulated in Ref. \cite{tamm15} and (b) equilibrium unknotted ring in a box simulated in Ref. \cite{imakaev14} (points) together with their best fits with Maxwell distribution \eq{maxwell} (lines). (c-d) Variances of the best fit Maxwell distributions as functions of $s$, (c) data from Ref. \cite{tamm15}, $N_e \approx 50$, (d) data from Ref. \cite{imakaev14}, $N_e \approx 150$.}
\label{fig:03}
\end{figure}

To check whether the distributions obtained in numeric simulations of topologically stabilized polymer states are Gaussian or not, we used the available numerical data from two independent sources: the conformations of a long unknotted ring in a box with reflecting boundary conditions studied in \cite{imakaev14}, and those of partially equilibrated crumpled globule conformations of linear chains with periodic boundary conditions generated in \cite{tamm15}. We plotted in \fig{fig:03} the distributions of monomer-to-monomer distance $\ve x \equiv |\ve x_k -\ve x_m|$ for different values of $s = |k-m|$ taken from the simulation data, and their best fit by the Maxwell distributions
\be
P(\ve x) = 4\pi \ve x^2 \left(\frac{3}{2\pi \sigma^2(s)}\right)^{3/2} \exp\left(-3 \frac{\ve x^2}{2\sigma^2(s)}\right)
\label{maxwell}
\ee
As one can see, apart from the very small values of $s$ the fits are remarkably good. The $\sigma^2(s)$ dependencies (see \fig{fig:03}c,d) exhibit a change in their shape around $s \approx N_e$ from the behavior typical for ideal polymer chains in a melt to a slower growth at large $s$, which indicates the presence of unscreened topological interactions. Remarkably, this behavior is very similar to that shown in \fig{fig:02}b.

We conclude, therefore, that the simple quadratic Hamiltonian \eq{quadr_a} with coefficients calibrated to match experimentally observed fractal dimension seems to be a very good candidate for effective phenomenological description of these states. Hopefully, further research will shed more light on which particular properties of topologically stabilized states (return probability, knot invariants, etc.) can be reproduced in this simple way and which need a more sophisticated approach. In any case, it seems clear that simple and exactly solvable phenomenological approach presented here would be a useful addition to the toolkit used for the study of this fascinating polymer states.

\begin{acknowledgments}
We are very grateful to M. Imakaev and A. Gavrilov who kindly provided us with the raw simulation data from refs. \cite{imakaev14} and \cite{tamm15}, respectively, to D. Grebenkov, R. Metzler, and G. Oshanin for numerous illuminating discussions and to A.Yu. Grosberg for critical comments on the manuscript. This work was supported by the EU-FP7-PEOPLE-IRSES grant DIONICOS (612707). SN is grateful to the RFBR grant 16-02-00252A for partial support, KP and MT acknowledge the support of the Foundation for the Support of Theoretical Physics and Mathematics ``BASIS'' (grant 17-12-278). Significant part of the work presented here was done during KP and MT visits to LPTMS at Universite Paris Sud, KP visits to the Theoretical Physics group at Potsdam University, and MT visits to Applied Mathematics Research Center at Coventry University. We use this opportunity to thank the hosts for their warm hospitality.
\end{acknowledgments}

\begin{appendix}
\section{Spectrum of the interaction matrix}

Consider a chain with Hamiltonian \eq{quadr_a} and coefficients behaving as $a(s) = c s^{-\gamma}$. Here we analyze the spectrum \eq{eigv} of the matrix $\ve A$ for the physical range of exponents, $\gamma > 2$. In the continuum limit one has:
\be
\omega_p = 2^{\gamma}c\,\left(\frac{N}{\pi p}\right)^{1-\gamma} \int_{2\pi p/ N}^{\pi p} x^{-\gamma} \left(1-\cos(x)\right)dx
\approx 2^{\gamma} c\,\left(\frac{N}{\pi p}\right)^{1-\gamma} \left\{ I(\gamma, p/N) - o\left(p^{1-\gamma}\right) \right\}
\ee
where the integral $I$ is
\be
I(\gamma, p/N) = \int_{2\pi p/ N}^{\infty}x^{-\gamma}\left(1-\cos(x)\right)dx =
\frac{1}{\gamma-1}\left(\frac{2\pi p}{N}\right)^{1-\gamma} - \Re\left[i^{\gamma-1}\; \Gamma\left(1 - \gamma, \frac{2\pi p i}{N} \right)\right]
\label{ig}
\ee
and $\Gamma$ is the holomorphic continuation of the upper incomplete $\Gamma$-function:
\be
\Gamma(s, z) = \Gamma(s) - \Gamma(s) z^s \exp(-z) \sum_{k=0}^{\infty} \frac{z^k}{\Gamma(s+k+1)} =
\Gamma(s) - \frac{1}{s} z^s + \frac{1}{s+1} z^{s+1} - \frac{1}{s+2} z^{s+2} + o\left(z^{s+2}\right)
\label{tay}
\ee
Using series \eq{tay} one can rewrite the real part in \eq{ig} as follows:
\be
\Re\left[i^{\gamma-1}\; \Gamma\left(1 - \gamma, \frac{2\pi p i}{N} \right)\right] =
\Gamma(1-\gamma) \cos\frac{\pi(\gamma-1)}{2} + \frac{1}{\gamma-1} \left(\frac{N}{2 \pi p}\right)^{\gamma-1} +
\frac{1}{3-\gamma} \left(\frac{2 \pi p}{N}\right)^{3-\gamma} + o\left(\left(\frac{\pi p}{N}\right)^{4-\gamma}\right)
\label{tay1}
\ee
Collecting \eq{tay1} and \eq{ig}, one ends up with the spectrum
\be
\frac{\omega_p}{c} = -2^\gamma\left(\frac{\pi p}{N}\right)^{\gamma-1}\Gamma(1-\gamma)\cos\frac{\pi(\gamma-1)}{2} -
\frac{8}{3-\gamma}\left(\frac{\pi p}{N}\right)^2 + o\left(\left(\frac{\pi p}{N}\right)^{3}\right)
\ee
which yields the following asymptotic in the limit $p/N \to 0$:
\be
\omega_p \sim \left\{ \begin{array}{cl} \disp &2^\gamma \; \Gamma(1-\gamma)
\cos\frac{\pi(3-\gamma)}{2}\left(\frac{\pi p}{N}\right)^{\gamma-1} \; \; \mbox{for $2 < \gamma < 3$} \medskip \\
\disp &\frac{8}{\gamma-3}\left(\frac{\pi p}{N}\right)^{2} \;\; \mbox{for $\gamma > 3$}
\end{array} \right.
\label{a6}
\ee

\end{appendix}

\end{document}